\documentclass[useAMS,usenatbib]{mnras}

\usepackage{graphicx}
\usepackage{ulem}
\usepackage{float}
\usepackage{psfig}
\usepackage{subcaption}
\usepackage{aecompl}
\usepackage[T1]{fontenc}

\title[Magnetic field of Arp 269]
{Probing the magnetic field of the nearby galaxy pair Arp 269}
\author[B. Nikiel--Wroczy\'nski et al.]
{
B. Nikiel-Wroczy\'nski\thanks{E-mail:iwan@oa.uj.edu.pl},
M. Jamrozy,
M. Soida,
M. Urbanik, and
J. Knapik\\
Astronomical Observatory, Jagiellonian
University, ul. Orla 171, Krak\'ow PL 30-244, Poland
}

\begin{document}

\date{Accepted xxxx. Received xxxx; in original form xxxx}

\pagerange{\pageref{firstpage}--\pageref{lastpage}} \pubyear{xxxx}

\maketitle

\label{firstpage}

\begin{abstract}

We present a multiwavelength radio study of the nearby galaxy pair Arp\,269 (NGC\,4490\slash 85). High sensitivity to extended structures
gained by using the merged interferometric and single dish maps allowed us to reveal a previously undiscovered extension of 
the radio continuum emission. Its direction is significantly different from that of the neutral gas tail, suggesting that  
different physical processes might be involved in their creation. The population of radio-emitting electrons is generally young, signifying an 
ongoing, vigorous star formation -- this claim is supported by strong magnetic fields (over 20\,$\mu$G), similar to the ones found in much larger 
spiral galaxies. From the study of the spectral energy distribution we conclude that the electron population in the intergalactic bridge between
member galaxies originates from the disk areas, and therefore its age (app. 3.7--16.9 Myrs, depending on the model used) reflects the
timescale of the interaction. We have also discovered an angularly near Compact Steep Source -- which is a member of a different
galaxy pair -- at a redshift of approximately 0.125.

\end{abstract}

\maketitle

\section{Introduction}
\label{intro}

Among the diversity of forms that intergalactic structures can adopt,
giant gaseous tails and streams accompanying galaxy interactions 
constitute one of the most impressive phenomena. Widespread, dense and wide,
they are able to transport matter tens of kiloparsecs away from the
parent objects. These tails are sometimes accompanied by streams of
stars, either formed in the tail, or inherited from progenitor galaxies.
Systems that contain such tails are usually referred to as the ``gas
streamers''. Notable examples include the Leo Triplet, where the tail
extends by more than 140\,kpc from NGC\,3628 \citep{haynes,hinew}, or
Arp\,143, where close encounter of the pair members results in mass loss
experienced by  one of the galaxies, while the tail traces its
trajectory around the companion \citep{tail}.\\

An open question concerning these intergalactic structures is the existence
and importance of magnetic fields inside them.
The belief that magnetic fields can be associated with such tails 
is plausible because of the ambipolar diffusion. In this process, neutral particles 
are transferred together with conducting plasma, to which they are collisionally coupled
to (see eg. \citealt{MS}, or \citealt{MP} and references therein). Therefore, 
it might be expected that the neutral gas tail can be accompanied by a magnetised outflow.
Ability to transport gas
far away from the intra-system medium gives a possibility that also the
magnetic field -- frozen into the gas -- can be  supplied to large
volumes of intergalactic space. 
Several examples of such objects are known
(eg. the Antennae, \citealt{antennae}); this raises an interesting question, 
whether these outflows and tidal tails can play a crucial role in the
magnetisation of cosmic space -- as it was suggested by \citet{kronberg}. 
Also, the dynamical importance of the
magnetic field in the evolution of such structures is yet unknown,
while it is known to be one of the major participants in dynamics
and evolution of galaxies itselves \citep{beckbook}. 
Outside the galaxies, both thermal gas and gravitational energy 
densities are lower than in the galactic disks, and therefore the fraction 
of the total energy attributed to the magnetic field is higher, than 
inside them.\\

Study of the
intergalactic structures yields a possibility to learn more about the
magnetic field itself.  Unfortunately, though several neutral gas
structures with magnetic field were found between the galaxies (eg.
\citealt{taffy}), there is
no definite observational evidence for magnetisation of the giant
outflows and tails.\\

Nearby galaxy pair NGC\,4490/85, the so-called ``Cocoon Galaxy'' is one
of the bright, nearby objects from the Shapley-Ames Catalogue
\citep{sa_revised}. The intergalactic space between the member galaxies is
filled with numerous areas of  star formation. This is a clear sign of
an ongoing interaction between the pair members. NGC\,4490 is a very special
case of an interacting system. Despite the small distance between member
galaxies of the pair and the  ``cocoon'' outlook, tidal forces have not
yet disrupted its disk, as the interaction is just starting. The aforementioned
stream of intergalactic star-forming regions between the galaxies, accompanied by a 
bridge of radio emission is the single hint of this process. Two other
phenomena that are usually explained on the basis of tidal interactions
-- giant neutral gas tails, and generally rapid star formation -- are
not triggered by the  merging process.
\citet{clemens_hi} suggested that NGC\,4490 is a very young galaxy that
rapidly forms stars at a constant rate and the giant tails result from
the wind caused by frequent supernovae explosions.\\

Given the high efficiency of the star formation in this system one can
conclude that strong magnetic fields should be present inside and outside of 
the galaxies. Arp\,269 is also interesting as it provides an unique 
possibility to study the issue of gas magnetisation in the eve of the 
interaction process. And as this galaxy pair is a dwarfish system, the
merging process might be similar to that which gave rise to the current
large galaxies -- agglutination of the smaller objects. Particularly 
interesting is the chance to survey the {\rm H}{\sc i} envelope for any
traces of radio continuum emission.\\

A detailed study of the radio emission of this galaxy pair has been
made by \citet{clemens}. These authors have  presented interferometric
maps of the emission at various radio frequencies. High resolution of
their images -- 12 arcseconds -- provided a detailed insight into the
morphology of the radio emission.  However, achieving such a small beam
requires usage of a loose configuration of the interferometer. This
leads to an incomplete sampling of the innermost part of the (u,v) plane,
hence, loss of the extended emission. This effect turns out to be important
in studies of NGC\,4490\slash 85 made by \citet{clemens}, who state
substantial flux density loss of 40\% at 8.46\,GHz. This precluded any search
for diffuse intergalactic structures.\\

An useful solution that allows to regain the missing flux density information
whilst retaining high resolution can be obtained  by merging the
interferometric and single dish data. The resultant image has high both
resolution and sensitivity to the extended structures. In a propitious
situation, nearly whole missing flux density and all structures can be
successfully restored (see eg. \citealt{3627}).\\

High frequency interferometric observations of intergalactic, diffuse 
emission suffer also from a limited primary beam size. Moreover, without
the continuous supply of relativistic particles, the radio spectrum of
intergalactic structures quickly steepens making the intergalactic 
emission undetectable at high frequencies. For this reason we performed
sensitive, low-frequency  observations using the Giant Metrewave Radio
Telescope (GMRT) at 0.61\,GHz. This enabled us to search for the  radio
emission produced by the low energy electrons that have longer lifetimes,
and their emission can be traced at larger distances from their supply
sources in galactic disks. As there were no previous detailed studies
of this galaxy that made use of the under-gigahertz frequencies, our 
observations enable more accurate estimations of the spectral
parameters, like the break frequency, or synchrotron age. Additionally, 
the large primary beam of GMRT permits to study structures of larger
angular scales than those observable at several GHz.\\

In our paper we present a new map at 0.61 GHz from our own GMRT 
observations together with the VLA maps at previously studied frequencies, 
merged with our Effelsberg data. This data are used to probe
the magnetic field in various regions within and outside the member
galaxies, allowing to study the radio emission of a very young galaxy in
the beginning of the merging process with unprecedented fidelity
and in a broad frequency range. 

\begin{figure*}
\resizebox{\hsize}{!}{\includegraphics{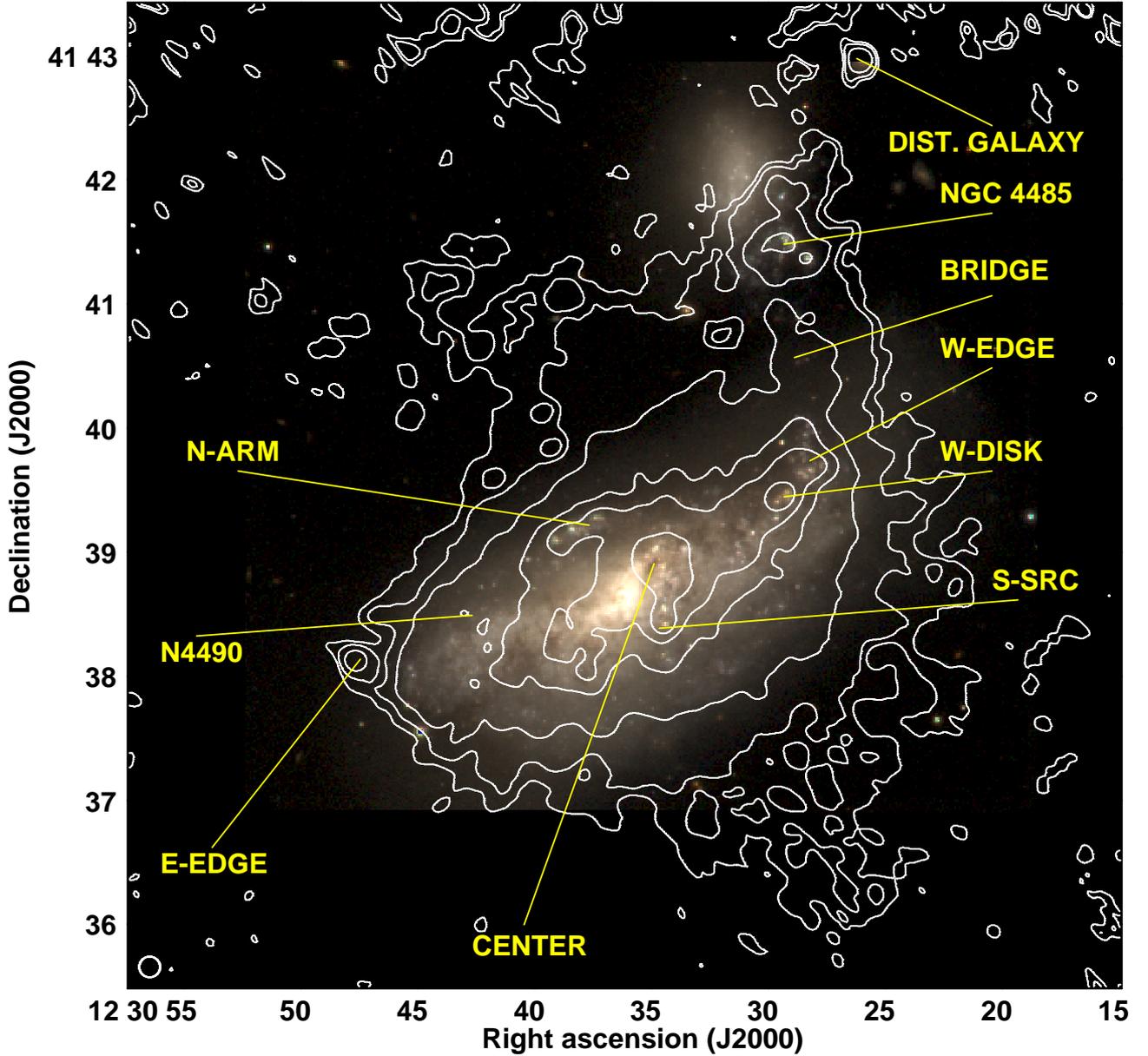}}
\caption{
Merged VLA + Effelsberg  map of the TP emission at 4.86\,GHz overlaid upon 
an RGB image, with positions of the regions discussed in this paper indicated.
The contour levels are $3,5,10,25,50,100 \times$ 33\,$\mu$Jy/beam
(first contour at 2.5 times the r.m.s. noise level). The angular resolution is 10\,arcsec.
The beam is represented by a circle in the lower left corner of the image.
Details of the maps used to produce this RGB composite can be found in the text
(Sect.~\ref{result}).
}
\label{RGB}
\end{figure*}

\section{Observations and data reduction}
\label{observ}

\subsection{GMRT data}
\label{obsgmrt}

The lowest frequency used in our study is 0.61\,GHz. These data have been obtained using the GMRT interferometer.
Observations were carried out in February, 2013.
The total bandwidth was 32\,MHz (divided into 256\,channels) and the total time-on-source (TOS) was 5.5\,h (integration 
time of 16\,s). Calibration, flagging and further reduction of these data was carried out using the Astronomical Image 
Processing System (\textsc{aips}). We used 3C\,286 to calibrate amplitudes (and bandpass), and the angularly close source 1227+365 to 
calibrate phases. A set of 49 sub-images (u,v)-tapered to obtain a circular beam has been made, and all of them
have been processed by a self-calibration pipeline to correct the phase information. The final map has been made
by the means of tessellation of the sub-images. Primary beam correction has also been applied.

\begin{table*}
\caption{\label{vladata}Basic information on the interferometric datasets used in this study. In case of multiple 
configuration datasets, the time-on-source (TOS) and noise values are given for the final ones. The angular 
resolution is 10\,arcseconds.}
\begin{center}
\begin{tabular}{rrrrrl}
\hline
\hline
Freq. & Telescope & Proj. code & Date &  TOS  & r.m.s. \\
 GHz  &		  &	       &      & [min] & [$\mu$Jy\slash 10'']\\
\hline
 0.61 & GMRT  & 23\_025 & 10.02.2013 & 335     & 60\\
 1.49 & VLA B & AA 116  & 21.08.1990 & 109     & 55\\
      & VLA C & AA 181  & 18.10.1994 &         &   \\
      & VLA D &		& 15.03.1995 &         &   \\
 4.86 & VLA C & AA 181  & 18.10.1994 &  86     & 40\\
      & VLA D &		& 15.03.1995 &         &   \\
 8.44 & VLA B & AA 181  & 26.12.1995 & 147     & 55\\
      & VLA C &		& 21.08.1990 &         &   \\
      & VLA D &		& 16.03.1995 &         &   \\
14.94 & VLA B & AJ 115  & 16.04.1985 & 132     &100\\
22.46 & VLA C & AJ 291  & 15.11.2002 & 215     & 50\\
\hline
\end{tabular}
\end{center}
\end{table*}

\subsection{Archive VLA data}
\label{obsvla}

The interferometric data at the frequencies higher than 1\,GHz have been taken from the NRAO Data Archive. We acquired data at 1.49, 
4.86, 8.44, 14.94 and 22.46\,GHz, recorded in different configurations and time windows. An overview of the datasets
used can be found in Table~\ref{vladata}. All these data have been calibrated following the standard procedure 
outlined in the {\sc aips} Cookbook.
After the initial analysis, we decided not to use the 14.94\,GHz data due to the low signal-to-noise ratio and small 
primary beam size. The 22.46\,GHz data are of better quality, but the primary beam area is even smaller and therefore
only the very center of this galaxy system is visible.
Concerning these issues, in order to broaden the frequency coverage of our study we used also the Ryle Telescope 
15.21\,GHz map from \citet{clemens}, courtesy of M. Clemens (details on the observational setup can also be found therein).

\subsection{Effelsberg data}

The single-dish total power data at 4.86 and 8.35\,GHz have been obtained in 2005, using the Effelsberg radio telescope 
\footnote{Based on observations with the \mbox{100m} telescope of the MPIfR (Max-Planck-Institut f\"ur Radioastronomie) at 
Effelsberg}. The lower frequency observations consisted of 32 coverages scanned in azimuth-elevation frame, whereas the higher 
frequency ones were made from 35 coverages recorded in R.A. and Declination. Both datasets were calibrated using 3C\,286 as the flux density 
calibrator (flux density values as given by \citealt{baars}) and reduced using the {\sc NOD2} package \citep{nod2}. Further details on 
how these maps have been processed will be presented in a forthcoming paper on the polarised emission from this galaxy pair 
(Knapik et al.~2016, in preparation).

\subsection{Merging procedure}

As mentioned in Sect.~\ref{intro}, one of the goals of our study was to achieve a modest resolution while having most of 
the flux density retained. In order to do so, merging procedure has been performed. We used {\sc aips} task {\sc imerg} and
followed an approach previously used eg. by \citet{3627}. To fulfill the procedure requirements, the single dish maps were
multiplied by the VLA primary beam at a given frequency and the image dimensions were transformed into a power of 2 (FFT requirement). 
The interferometric data have been transformed using the task {\sc hgeom} to make the geometry of the high- and low-resolution 
images consistent. The (u,v)-range of overlapment was chosen basing on the task's internal calculation of the normalisation 
factor (which should be equal to the ratio of the beam areas of the images merged). The resulting maps have been corrected 
for the primary beam. As a consistency check, we measured the total flux density in the maps before and after merging.
At 4.86\,GHz, the values for the final map agree within the 5 per cent error with the single-dish data. At 8.44\,GHz, the total flux density
of NGC\,4490 in the merged map is lower; this is mostly because of the negative 'bowl' that surrounds the galactic disk in the purely
interferometric map (an effect of sparse sampling of the (u,v)-plane due to the short time of observation). For both of 
these frequencies, flux densities of the point sources before and after merging are consistent. Therefore, we consider that the merging procedure
went out with a success.\\

For the sake of strictness we note here, that the central frequencies of the used X-band (8\,GHz) VLA and Effelsberg maps differ 
insensibly; the flux density difference does not exceed 1 per cent for the radio sources with spectral indices not steeper than
1 (using the $S_{\nu}\propto\nu^{-\alpha}$ definition of the spectral index $\alpha$) -- far lower than the assumed calibration 
errors. A detailed description of the merging process with the theoretical basis explained can be find in \citet{merging}.

\subsection{Error estimates}

To include the calibration errors, we decided to adopt 8 per cent uncertainty for the GMRT data, and 5 per cent for both VLA and
merged VLA+Effelsberg data. For the extended sources, a noise term associated with the size (product of the map noise and square root 
of the number of beams per structure) was introduced, and the final error is a sum of the calibration and noise errors.
For the non-thermal flux density (see Sect.~\ref{separation} for details), the error estimate contains also a 10 per cent uncertainty
of the thermal flux density estimation, that has been added to the final error estimate.

\section{Results}
\label{result}

To familiarise the readers with different structures described and analysed in this paper, we include an explanatory image 
of the radio emission with designations of individual sources indicated (Fig.~\ref{RGB}). Contours of the radio emission from merged 
(VLA+Effelsberg) radio data at 4.86\,GHz have been overlaid upon a composite RGB image made from the Johnson's RVB maps produced 
by the Vatican Advanced Technology Telescope \citep{vatt}. Details on the characteristics of the emission of these entities can
be found in Sect.~\ref{sources}.\\

\begin{figure*}
    \begin{subfigure}[p]{0.5\textwidth}
    \includegraphics[width=\textwidth]{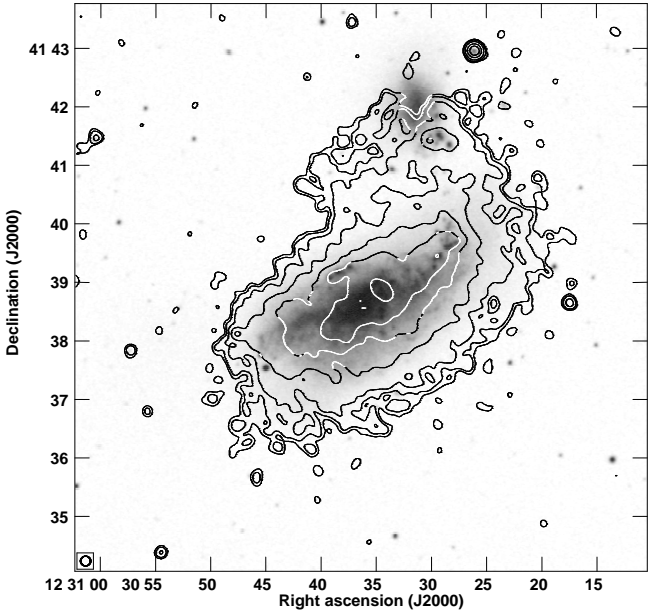}
    \captionsetup{width=0.9\textwidth}
    \caption{
      GMRT map at 0.61\,GHz.  
      The r.m.s. noise level is\\ 60\,$\mu$Jy/beam  
    }
    \label{gband}
    \end{subfigure}%
    \begin{subfigure}[p]{0.5\textwidth}
    \includegraphics[width=\textwidth]{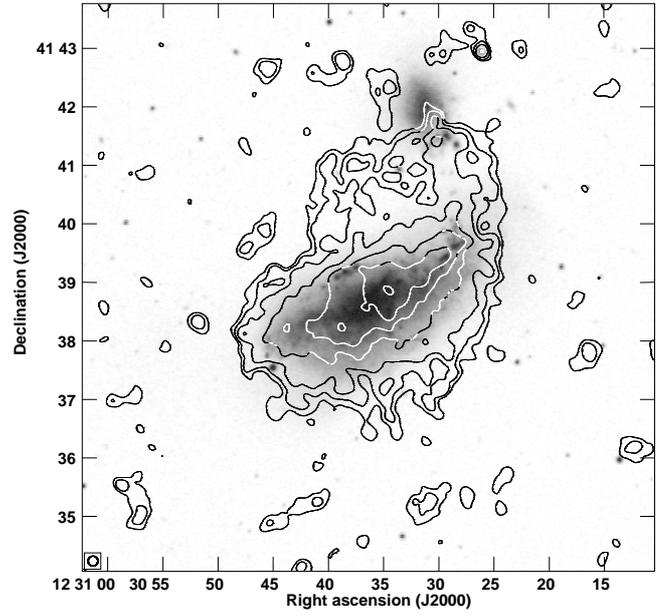}
    \captionsetup{width=0.9\textwidth}
    \caption{
      VLA map at 1.49\,GHz.  
      The r.m.s. noise level is 55\,$\mu$Jy/beam. 
    }
    \label{lband}
    \end{subfigure}%

    \begin{subfigure}[p]{0.5\textwidth}
    \includegraphics[width=\textwidth]{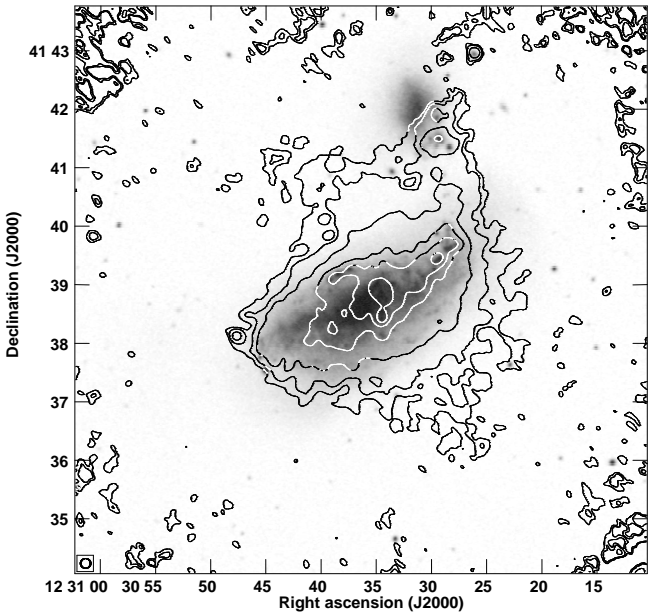}
    \captionsetup{width=0.9\textwidth}
    \caption{
      Effelsberg+VLA map at 4.86\,GHz.  
      The r.m.s. noise level is 40$\mu$Jy/beam.   
    }
    \label{cband}
    \end{subfigure}%
    \begin{subfigure}[p]{0.5\textwidth}
    \includegraphics[width=\textwidth]{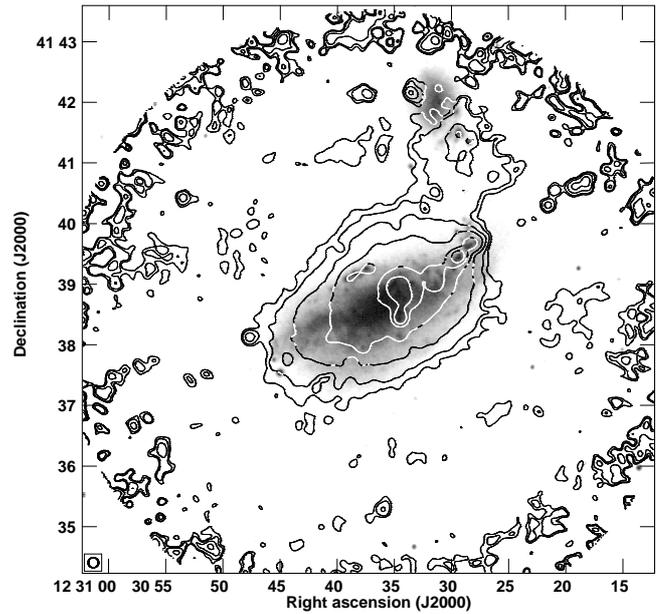}
    \captionsetup{width=0.9\textwidth}
    \caption{
      Effelsberg+VLA map at 8.44\,GHz.  
      The r.m.s. noise level is 55$\mu$Jy/beam. 
    }
      \label{xband}
     \end{subfigure}%
    \caption{
	  Maps of the TP emission of NGC\,4490 overlaid upon a POSS-II R-band image.
	  The contour levels are $3,5,10,25,50,100,250\times$ r.m.s. noise level. 
	  The 10-arcsec beam is represented by a circle in the lower left corner of the image.
	}
\end{figure*}

The GMRT map at 0.61\,GHz (Fig.~\ref{gband}) constitutes the lowest frequency image in our study. The radio emission from the 
galaxy pair is fairly strong, emerging not only from the disk of NGC\,4490, but also from several sources associated with
NGC\,4485, and from an intergalactic bridge that connects members of this galaxy pair. The northwestern side of NGC\,4490 is 
more pronounced than the opposite one. North from the pair, a point-like source can be seen. This source is a distant galaxy,
not related to Arp\,269.
It is clearly visible that the emission in the GMRT map is less extended in the southern direction than in the high-frequency 
map at 4.86\,GHz. At the same time, the northern part reaches its maximum extension at 0.61\,GHz. This is because our GMRT data
lack short spacing information; the (u,v)-plane coverage is moreover comparable to the VLA B\slash C configuration at L-band 
(1.49\,GHz). This results in a lower sensitivity to extended structures and hence only the northwestern extension, which is 
bright enough, is visible. Sharp edges of the emission -- manifesting as an overdensity of the 3 and 5 $\sigma$ contours
-- confirm this statement.\\

The 1.49\,GHz data (Fig.~\ref{lband}) also come from the interferometer only. Merged B, C, and D configurations of the VLA 
provide (u,v)-coverage dense enough in the innermost part of the plane to recover emission extended similarly to that at 0.61\,GHz. Not 
only does the disk extend in the NW direction, but also a weaker, albeit still easily detectable extension can be found on the
SE side. Intergalactic bridge and distant galaxy can also be seen.\\

The merged 4.86\,GHz data (Fig.~\ref{cband}) combine high resolution of the VLA with an excellent sensitivity to extended 
emission provided by the single dish data. Also, the primary beam is reasonably large, so that the regions relevant to our study do
not fall into its outskirts, where flux density uncertainties are higher. Worth noticing is the extent of the intergalactic emission,
which at 4.86\,GHz spans widely in both NE and SW directions. A detailed comparison of this structure to the giant {\rm H}{\sc i} 
tails is presented in Sect.~\ref{hisect}.\\
The intergalactic bridge between NGC\,4490 and 4485
is still visible, but less prominent than at lower frequencies. The distant radio galaxy lies close to the edge of the 
primary beam area, but still is easily detectable.\\

At 8.44\,GHz (Fig.~\ref{xband}), large extensions of the radio 
emitting area vanish. Only a small remnant of the northern structure can be seen; the southern one is completely undetected.
The galactic disk is visibly shrunk, especially in the western direction, and its optical boundaries lie within the outermost areas
of the lowest radio contour. The intergalactic bridge is still visible; however, as the primary beam is rather small at this frequency,
the companion galaxy lies at its very edge. The total recovered flux density (175 $\pm$ 9 mJy) is slightly lower that the single dish
amount (app. 200 mJy, as stated by \citealt{clemens}). This might be either an effect of the primary beam correction, or negative 
signal in the interefometric map (due to the sparse sampling of the (u,v)-plane).\\

At two highest frequencies, 14.94 and 22.46\,GHz, the primary beam is so small that it does not contain the whole galaxy. Also,
the sampling of the (u,v)-plane becomes more sparse and thus the sensitivity to extended structures drops significantly. In 
both of these maps, only the brightest, compact sources are visible. One of them is the point-like source south from the 
galactic centre; because of the low quality of the 14.94\,GHz map, more emission surrounding it is detected at 22.46\,GHz -- 
there is a short ridge of extended emission elongated in the N--S direction. The second source that is visible on both of these
maps is the western disk region of star formation (W--DISK).

\section{Analysis and discussion} 
\label{discus}

In this section we describe the procedure used to separate the thermal and non-thermal emission contents 
(Sect.~\ref{separation}) and results of estimation of parameters for which it is required: non-thermal spectral 
index (Sect.~\ref{spixtext}), total magnetic field (Sect.~\ref{magfield}), and spectral age of the electron population 
(Sect.~\ref{age}). Particular regions are described in Sect.~\ref{sources}. Coincidence between the continuum structures
that extend north and south from NGC\,4490 and the neutral hydrogen distribution is also discussed (Sect.~\ref{hisect}). 
Remarks on the (non-detected) remnant of SN2008ax and the detected distant galaxy can be found in Sect.~\ref{snr} and
\ref{distgal}, respectively. Values
measured and derived can be found in two tables: Table~\ref{flux densityess} lists the total and non-thermal flux densityes of the selected
structures, while Table~\ref{params} contains information regarding the magnetic field, injection spectral index, break
frequency and spectral age of these entities. Location of the specific sources have been marked in Fig.~\ref{RGB}.

\begin{table*}
\caption{Flux densities (with errors) obtained for the selected regions of radio emission.}
\begin{center}
\begin{tabular}[]{llcccccc}
Region	& Component	& \multicolumn{6}{c}{Flux density in mJy at a given frequency, expressed in GHz} \\
\multicolumn{2}{c}{} & 0.61 & 1.49 & 4.86 & 8.44 & 15.21 & 22.46 \\	
\hline
Total & TOT & 1426 $\pm$ 116 & 800 $\pm$ 41 & 357 $\pm$ 19 & 175 $\pm$ 9 & 95 $\pm$ 5 & 
$^{*}$ \\
      & NTH & 1386 $\pm$ 120 & 761 $\pm$ 45 & 322 $\pm$ 22 & 139 $\pm$ 13 & 69 $\pm$ 8 &
$^{*}$ \\
Smooth Disk & TOT & 1360 $\pm$ 116 & 763 $\pm$ 41 & 328 $\pm$ 19 & 155 $\pm$ 9 & 74 $\pm$ 5 &
$^{*}$ \\
      & NTH & 1336 $\pm$ 120 & 733 $\pm$ 45 & 302 $\pm$ 22 & 127 $\pm$ 13 & 56 $\pm$ 8 &
$^{*}$ \\
NGC\,4485 & TOT & 2.80 $\pm$ 0.24 & 1.85 $\pm$ 0.15 & 0.90 $\pm$ 0.10 & $^{*}$ & 0.94 $\pm$ 0.11 & 
$^{*}$ \\
      & NTH & 2.45 $\pm$ 0.25 & 1.36 $\pm$ 0.18 & 0.46 $\pm$ 0.12 & $^{*}$ & 0.42 $\pm$ 0.15 &
$^{*}$ \\
Bridge & TOT & 27.82 $\pm$ 1.66 & 15.43 $\pm$ 0.98 & 6.54 $\pm$ 0.51 & 3.82$ \pm$ 0.31  & 1.34 $\pm$ 0.25 & 
$^{*}$ \\
       & NTH & 27.39 $\pm$ 1.71 & 15.04 $\pm$ 1.02 & 6.19 $\pm$ 0.54 & 3.42$ \pm$ 0.27 & 1.03 $\pm$ 0.28 &
$^{*}$ \\
N-Arm & TOT & 7.80 $\pm$ 0.58 & 4.50 $\pm$ 0.32 & 3.31 $\pm$ 0.23 & 2.50 $\pm$ 0.22 & 1.41 $\pm$ 0.16 & 
$^{*}$ \\
      & NTH & 5.43 $\pm$ 0.71 & 3.20 $\pm$ 0.40 & 1.69 $\pm$ 0.32 & 0.99 $\pm$ 0.31 & 0.46 $\pm$ 0.21 &
$^{*}$ \\
Center & TOT & 28.85 $\pm$ 2.55 & 16.37 $\pm$ 1.04 & 12.03 $\pm$ 0.77 & 8.90 $\pm$ 0.61 & 7.00 $\pm$ 0.53 & 
1.24 $\pm$ 0.18\\
      & NTH & 23.99 $\pm$ 3.04 & 11.66 $\pm$ 1.51 & 8.24 $\pm$ 1.14 & 5.09 $\pm$ 0.99 & 3.63 $\pm$ 0.86 &
0.04 $\pm$ 0.01\\
E-Edge & TOT & 2.25 $\pm$ 0.24 & 1.09 $\pm$ 0.11 & 0.46 $\pm$ 0.08 & 0.42 $\pm$ 0.06 & 0.27 $\pm$ 0.06 & 
$^{*}$ \\
      & NTH & 2.25 $\pm$ 0.24 & 1.16 $\pm$ 0.12 & 0.46 $\pm$ 0.08 & 0.43 $\pm$ 0.06 & 0.27 $\pm$ 0.06 &
$^{*}$ \\
W-Disk & TOT & 9.69 $\pm$ 0.88 & 6.07 $\pm$ 0.40 & 4.65 $\pm$ 0.30 & 4.02 $\pm$ 0.29 & 2.70 $\pm$ 0.21 & 
1.41 $\pm$ 0.12\\
      & NTH & 7.97 $\pm$ 1.05 & 4.59 $\pm$ 0.54 & 3.93 $\pm$ 0.37 & 2.87 $\pm$ 0.41 & 1.90 $\pm$ 0.29 &
0.86 $\pm$ 0.10\\
W-Edge & TOT & 10.34 $\pm$ 0.96 & 4.12 $\pm$ 0.33 & 3.43 $\pm$ 0.26 & 1.68 $\pm$ 0.21 & 1.77 $\pm$ 0.18 & 
$^{*}$ \\
      & NTH & 6.01 $\pm$ 1.39 & 2.83 $\pm$ 0.46 & 2.70 $\pm$ 0.33 & 1.28 $\pm$ 0.25 & 1.10 $\pm$ 0.25 &
$^{*}$ \\
Dist. Galaxy & TOT & 5.06 $\pm$ 0.31 & 1.47 $\pm$ 0.13 & 0.49 $\pm$ 0.06 & $^{*}$  & $^{*}$  & 
$^{*}$ \\
      & NTH & $^{**}$  & $^{**}$ & $^{**}$ & $^{**}$ & $^{**}$ & $^{**}$ \\
\hline
\hline
\end{tabular}
\end{center}
\begin{flushleft}
$^{*}$~Area not included into the primary beam\\
$^{**}$~~Area not covered by the H$_{\rm \alpha}$ map\\
\end{flushleft}
\label{flux densityess}
\end{table*}

\subsection{Thermal flux density separation}
\label{separation}

A crucial issue when describing the radio emission is a proper separation of the thermal and non-thermal components. 
There are several possible ways to deal with this problem. The spectrum can be described as
a sum of two power-law components: thermal (scaling with the power of 0.1) and synchrotron (scaling with
the parameter $\alpha_{\rm nth}$) ones. We attempted to do a pixel-by-pixel fit of the values from the radio maps.
However, it turns out that even broader range of frequencies is necessary, as the fitted values have shown 
strong dependence on the initial ones, yielding the results unreliable. Hence, we decided to perform a direct
subtraction of the thermal flux density. This can be done basing on the knowledge of the H$_{\alpha}$ flux density, which is closely 
related to the radio emission.\\

We have re-created the procedure described by \citet{heesen}. We took the map of the H$_{\alpha}$ emission from the Spitzer
Local Volume Legacy (LVL) database \citep{lvl}, which was calibrated using the standard procedure described in the LVL cookbook.
It was then re-scaled to more useful units to be used for the thermal 
flux density estimation using the conversion equation presented by \citet{ddb}, where the correction for galactic extinction was 
conducted using $E(B-V)=0.019$, as basing on \citet{sfd,sf}. The H$_{\alpha}$ flux density is a subject to the dust absorption
in the host galaxy; this effect can be corrected with the usage of the dust emission flux density multiplied by a specific coefficient
\citep{calzetti, evans}. We decided to use the Spitzer MIPS 24\,$\mu$m data, also available from the LVL archive.
It should be explicitly noted here, that despite very low (several per cents) foreground extinction, the unknown internal
one (inside the dense regions of radio emission) can still be significant (see Sect.~\ref{sources}). Therefore,
derived values for the thermal flux density should be assumed to be the lower boundaries for its real value; this, in turn, implies
that the non-thermal flux density values can be considered as its upper limit (however, one should bear in mind that the purely 
interferometric maps might be a subject to the zero spacing problem). {\sc aips} was used to subtract the thermal content in order 
to produce maps of the non-thermal emission. These maps 
were then used for the magnetic field and spectral parameters estimation.

\subsection{Spectral index}
\label{spixtext}

\begin{figure}
\resizebox{\hsize}{!}{\includegraphics{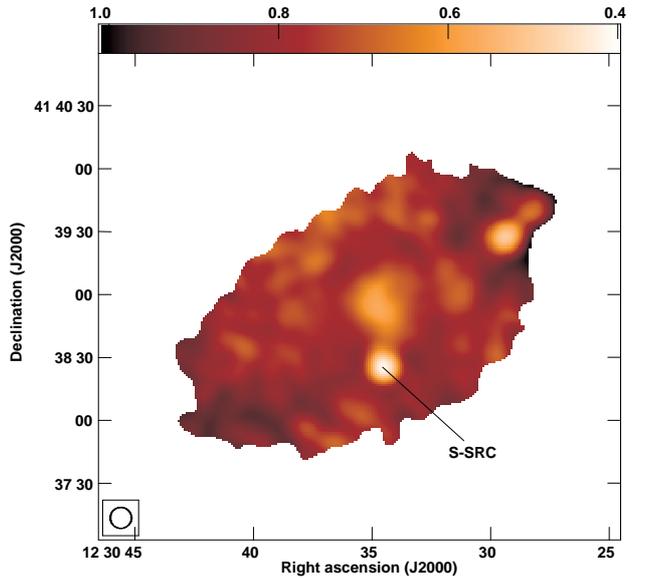}}
\caption{
 Map of the non-thermal spectral index calculated between 0.61 and 4.86\,GHz 
 with position of the S-SRC indicated}. The angular resolution is 10\,arcsec. 
\label{spixGC}
\end{figure}

The sole non-thermal emission yields a possibility to calculate the non-thermal spectral index. We decided to 
calculate a three-point spectral index map, using the GMRT 0.61\,GHz and the VLA+Effelsberg 4.86, and 8.44\,GHz data. 
These maps have the highest fidelity among our set, and provide a rather broad coverage of frequencies. 
All were trunctated at the 10 r.m.s. level to avoid generation of unreliable spectral index values. 
However, as the (u,v) coverages of these three maps are different, values at the edges of the spectral index distribution
should be treated with caution. Nevertheless, these derived for the star-forming regions and the galactic disk are not that much affected: 
compact character of the regions of star formation reduces importance of the shortest baselines, and for the galactic disk the 
radio emission is powerful enough that the losses due to reduced sensitivity for the largest and weakest structures are not significant.
The map of the spectral index can be seen in Fig.~\ref{spixGC}. Values for each of the pixels were calculated 
using a combination of \textsc{aips} tasks, following a general recipe:

 $$\alpha=-{N\sum\limits_{i=1}^{N}(\log S_i\log\nu_i)-\sum\limits_{i=1}^{N}\log S_i\sum\limits_{i=1}^{N}\log\nu_i\over
N\sum\limits_{i=1}^{N}\log\nu_i^2-\left(\sum\limits_{i=1}^{N}\log\nu_i\right)^2}$$

This equation is valid when using the $S_{\nu}\propto\nu^{-\alpha}$ convention for the spectral index, which we have adopted throughout
this paper.
The disk emission is characterised by indices of 0.6--0.75, typical for a non-thermal, not yet aged electron population. Several spots 
of flatter spectrum can be seen. The most notable one is S-SRC, for which the separation of the thermal and non-thermal content
was not entirely successful (see Sect.~\ref{sources} for details). Other areas correspond to the star-forming regions 
dispersed through the disk.\\

\subsection{Magnetic field}
\label{magfield}

Estimates of the strength and energy density of the magnetic field can be derived from the non-thermal intensity and 
non-thermal spectral index under the assumption of energy equipartition between the cosmic rays and the magnetic field. A set 
of necessary formulae was presented by \citet{bfeld}. These Authors have also released the code \textsc{bfeld}
that computes the magnetic field strength as well as its energy density. \textsc{bfeld} uses following parameters: total
pathlength through the source $D$, proton--to--electron energy density ratio $K_0$, non-thermal spectral index $\alpha$, and 
the mean synchrotron surface brightness of the chosen region. For $K_0$ we used the typical value of 100, which 
is right even for the starburst galaxies \citep{beckstarburst}.
It should be noted here, that $D$ is not very well known; this prevented us from creating a map of the magnetic field.
We have adopted values of 0.5 to 1.0\, kpc and taken info the account the inclination of galactic disks.
Results of the magnetic field estimation can be found in Table~\ref{params}.
The uncertainty of the total magnetic field accounts for factor two variations in $D$, $\alpha$, and the surface
brightness.

\begin{table*}
\caption{Parameters used and derived in magnetic field and spectral properties estimations. Explanation is provided in the 
text.}
\begin{center}
\begin{tabular}[]{lccccccccc}
Region	& $\alpha_{\rm NTH}$ & B$_{\rm TOT}$ & U$_{\rm B}$ & $\alpha_{\rm inj, JP}$ & $\alpha_{\rm inj, CI}$& 
$\nu_{\rm break,JP}$ & $\nu_{\rm break,CI}$ & Age (JP) & Age (CI)\\
	&   & [$\mu$G] & [$\mathrm{erg}\,\mathrm{cm^{-3}}$] &  &  & [GHz] & [GHz] & [Myr] & [Myr]\\
	\hline
Total		& 0.70  & 21.9 $\pm$ 2.9 & 1.91 $\pm$ 0.50 & 0.52 & 0.60 &  11 -- 27  & 1.5 -- 11 
& 2.4 -- 5.7  	& 3.8 -- 14.9 \\
Sm. Disk	& 0.71  & 21.5 $\pm$ 2.8 & 1.85 $\pm$ 0.48 & 0.48 & 0.63 & 8.5 -- 20  & 1.5 -- 11 
& 2.9 -- 6.5	& 3.9 -- 15.1 \\
NGC\,4485	& 0.80  & 15.5 $\pm$ 2.0 & 0.96 $\pm$ 0.24 & 0.71 & 0.71 &    $>$ 15  & $>$ 2 
& $<$ 7.8	& $<$ 23.5	\\
Bridge		& 0.72  & 18.3 $\pm$ 2.0 & 1.39 $\pm$ 0.28 & 0.45 & 0.57 &   9 -- 20  &   2 -- 14
& 3.7 -- 7.9	& 4.5 -- 16.9	\\
N-Arm		& 0.56  & 25.5 $\pm$ 3.5 & 2.61 $\pm$ 0.71 & 0.43 & 0.46 & 9.5 -- 629 &   1 -- 293
& 0.4 -- 4.9  	& 0.6 -- 15.1 \\
Center		& 0.51  & 39.3 $\pm$ 5.4 & 6.19 $\pm$ 1.71 & 0.45 & 0.46 &    $>$ 34  & $>$ 8 
& $<$ 1.4	& $<$ 2.8	\\
E-Edge		& 0.76  & 21.7 $\pm$ 2.8 & 1.89 $\pm$ 0.49 & 0.69 & 0.69 &    $>$ 59  & $>$ 16
& $<$ 2.4	& $<$ 4.7	\\ 
W-Disk		& 0.61  & 26.0 $\pm$ 3.5 & 2.72 $\pm$ 0.73 & 0.38 & 0.38 &    $>$ 54  & $>$ 16
& $<$ 2.0	& $<$ 3.7	\\
W-Edge		& 0.84  & 18.2 $\pm$ 2.3 & 1.33 $\pm$ 0.33 & 0.38 & 0.38 &    $>$ 26  & $>$  6
& $<$ 4.7	& $<$ 6.8\\
\hline
\end{tabular}
\end{center}
\label{params}
\end{table*}

\subsection{Age of the structures}
\label{age}

Successful separation of the thermal and non-thermal emission opens
a possibility to calculate the so-called spectral age, defined as 
the amount of time elapsed since the last acceleration
of the electrons that are responsible for the observed radiation. 
However, as various physical processes are characterised by different
time-scales of the electron spectrum decay, one can choose from 
several models of the electron losses. We decided to follow the 
procedure used by us in a previous paper \citep{our143}. Hence,
estimations of the break frequency were done using the \textsc{synage} 
package \citep{murgia}, which allows to test Jaffe--Perola 
(\citealt{JP}, JP), Kardashev--Pacholczyk 
(\citealt{kardashev};~\citealt{pacholczyk}, KP), or continuous injection 
(\citealt{pacholczyk};~\citealt{myears};~\citealt{carilli}, CI) models 
of electron energy losses. \textsc{synage} uses the spectral energy 
distribution (SED) to determine following parameters: spectral index of 
the injected electron population $\alpha_{\mathrm{inj}}$ and the 
frequency above which the observed spectrum steepens from the initial one
-- the spectral break frequency $\nu_{\mathrm{break}}$ (expressed in GHz).\\

\begin{figure*}
    \begin{subfigure}[p]{0.45\textwidth}
    \includegraphics[width=0.7\textwidth, angle=270]{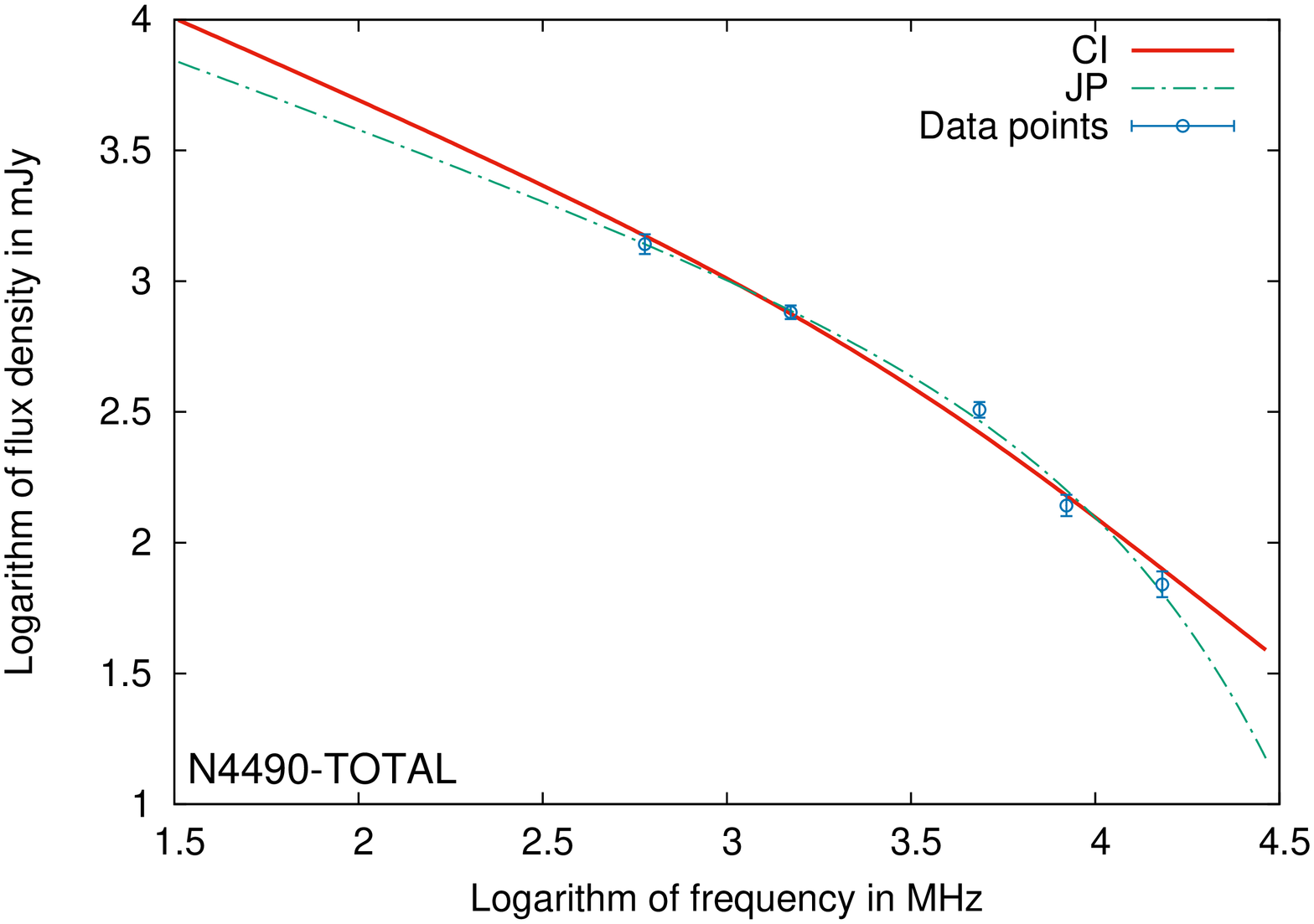}
    \end{subfigure}%
    \begin{subfigure}[p]{0.45\textwidth}
    \includegraphics[width=0.7\textwidth, angle=270]{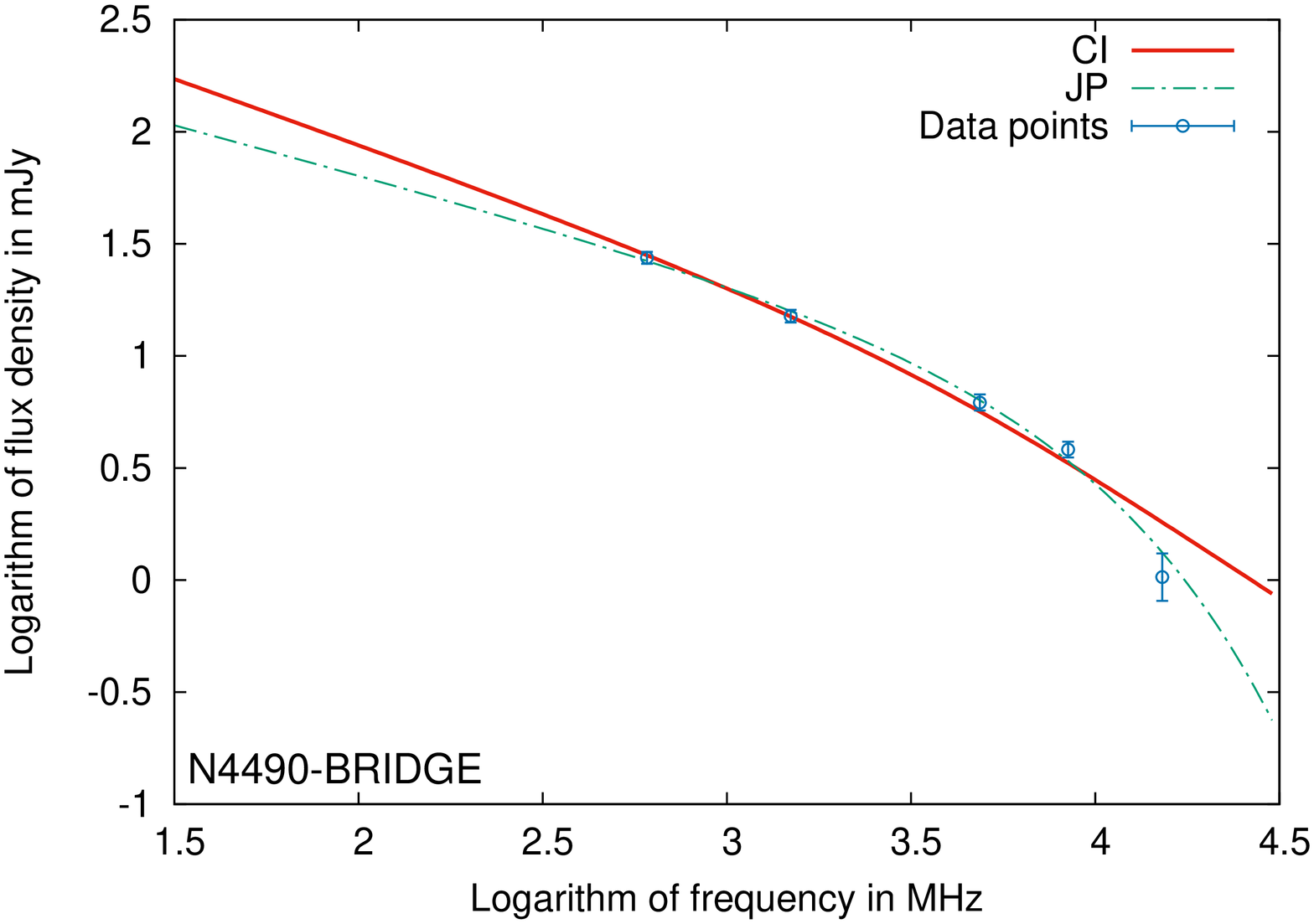}
    \end{subfigure}
    \begin{subfigure}[p]{0.45\textwidth}
    \includegraphics[width=0.7\textwidth, angle=270]{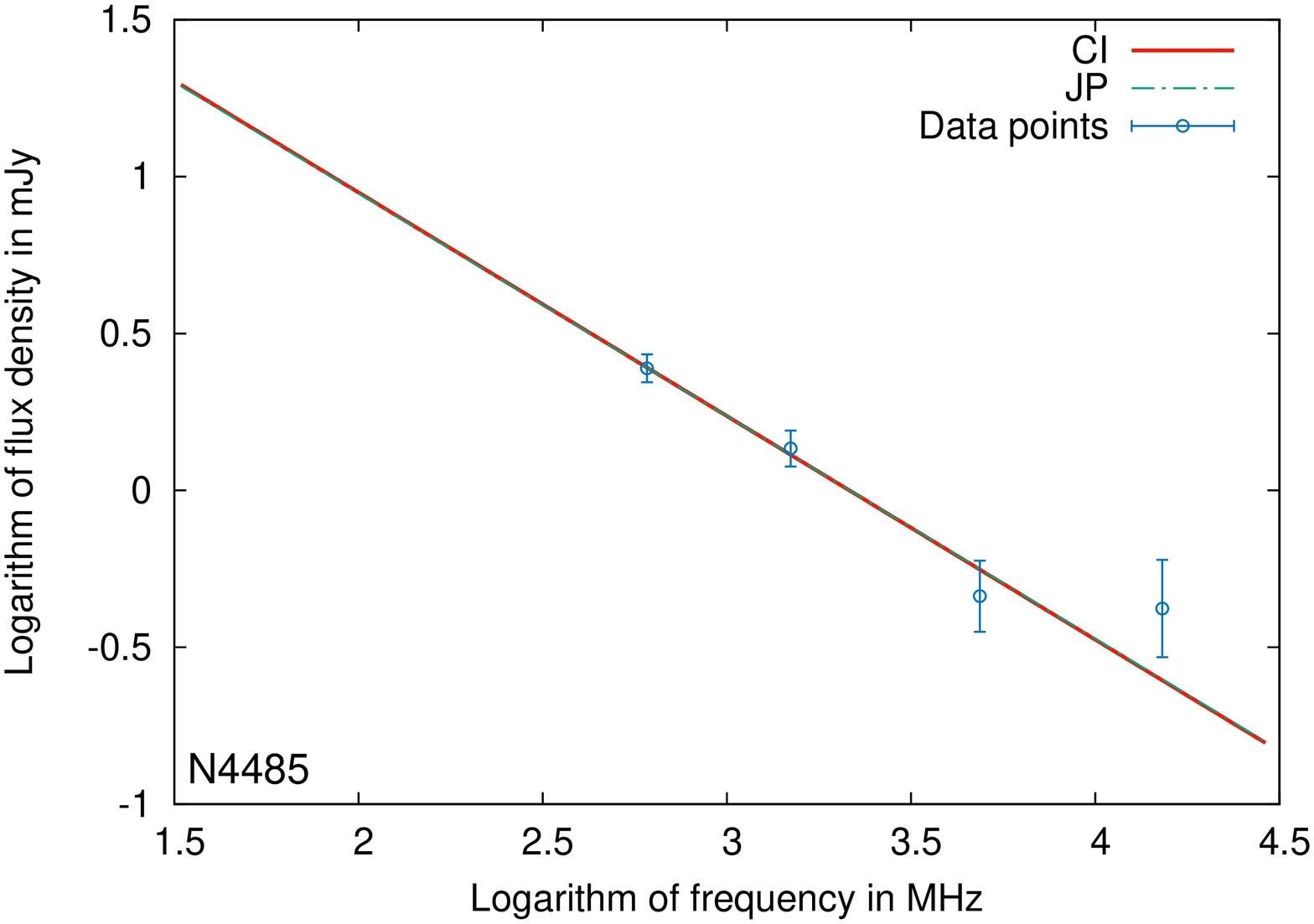}
    \end{subfigure}%
    \begin{subfigure}[p]{0.45\textwidth}
    \includegraphics[width=0.7\textwidth, angle=270]{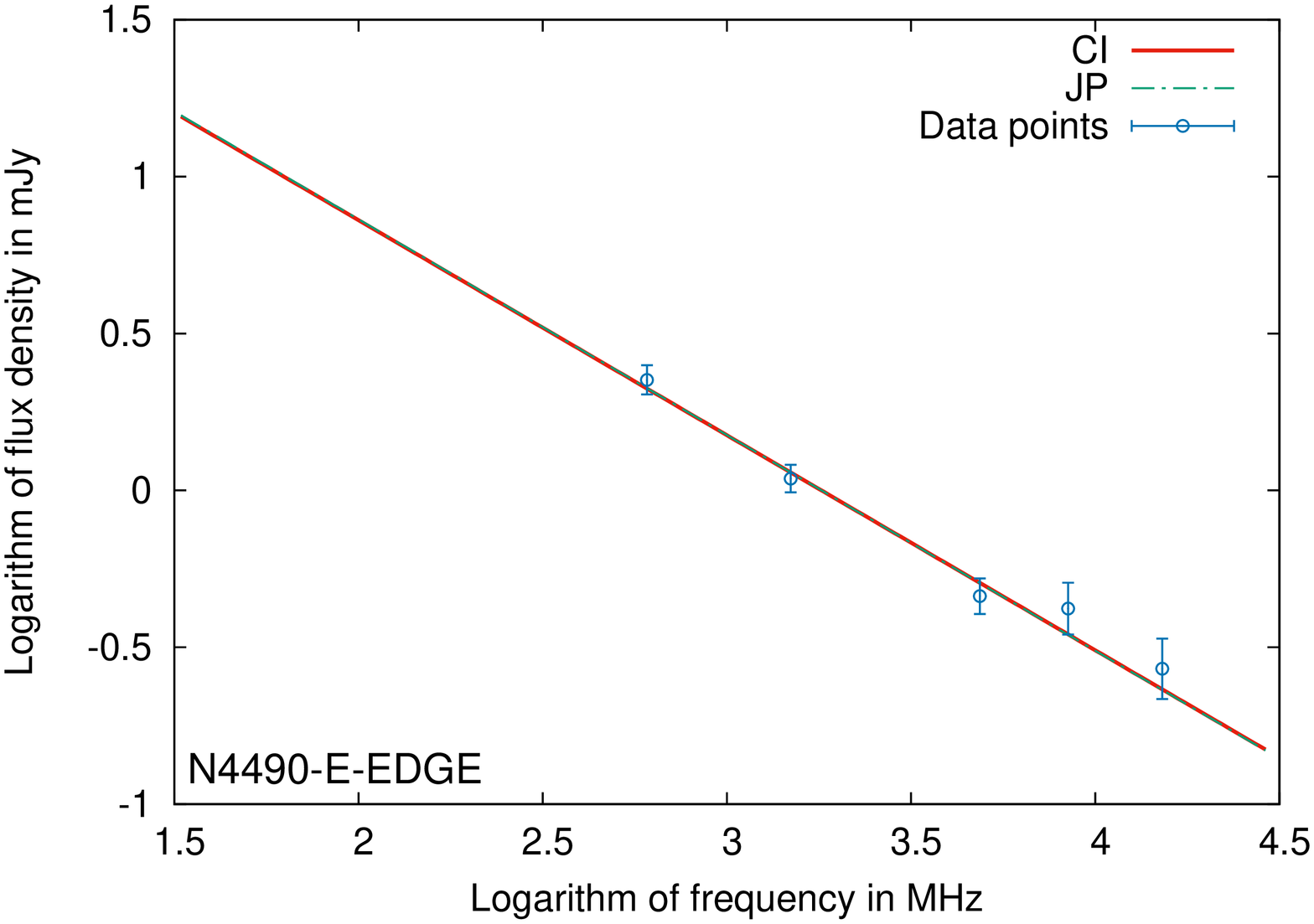}
     \end{subfigure}
         \begin{subfigure}[p]{0.45\textwidth}

    \includegraphics[width=0.7\textwidth, angle=270]{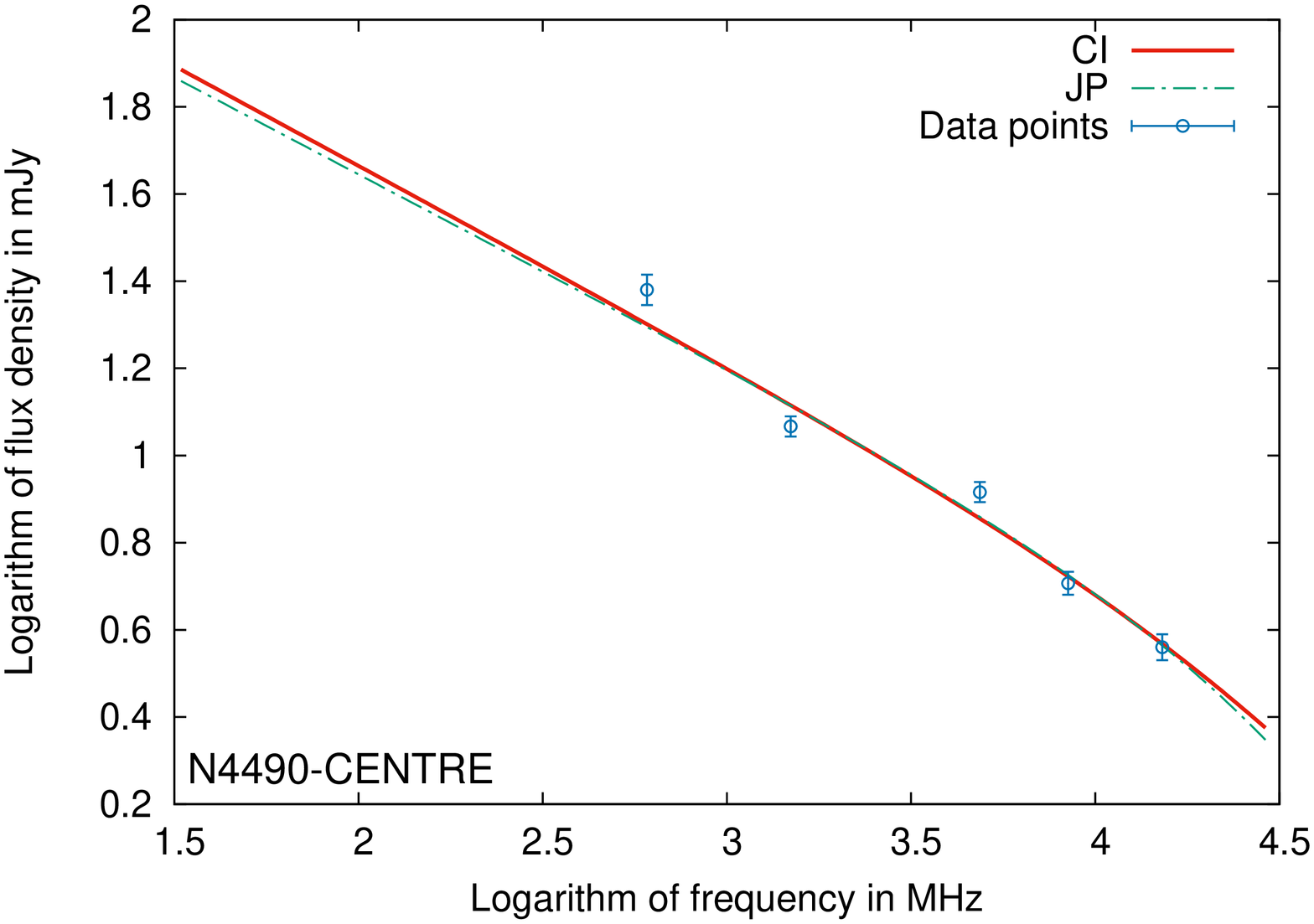}
      \end{subfigure}%
    \begin{subfigure}[p]{0.45\textwidth}
   \includegraphics[width=0.7\textwidth, angle=270]{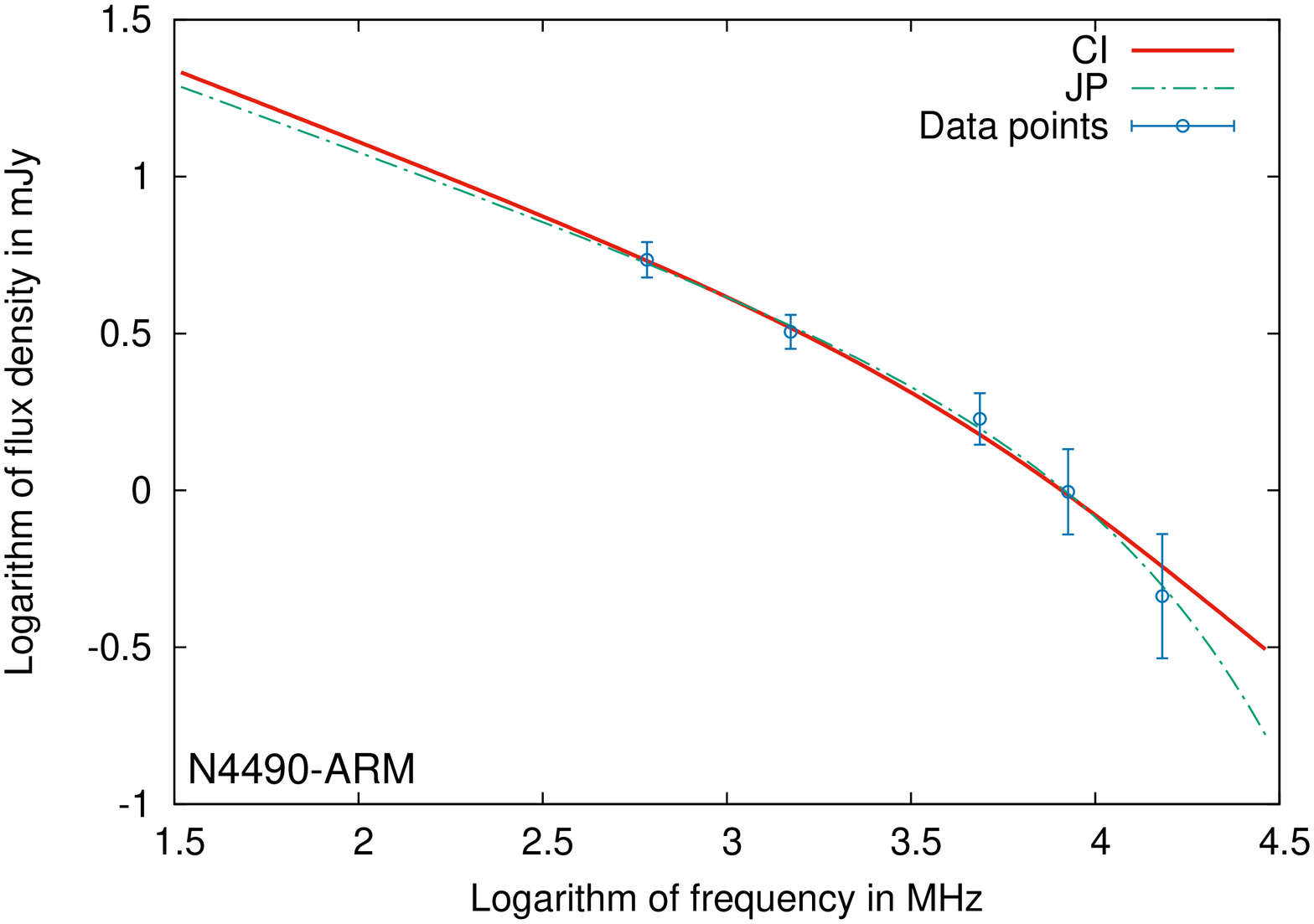}
      \end{subfigure}
      
    \begin{subfigure}[p]{0.45\textwidth}
   \includegraphics[width=0.7\textwidth, angle=270]{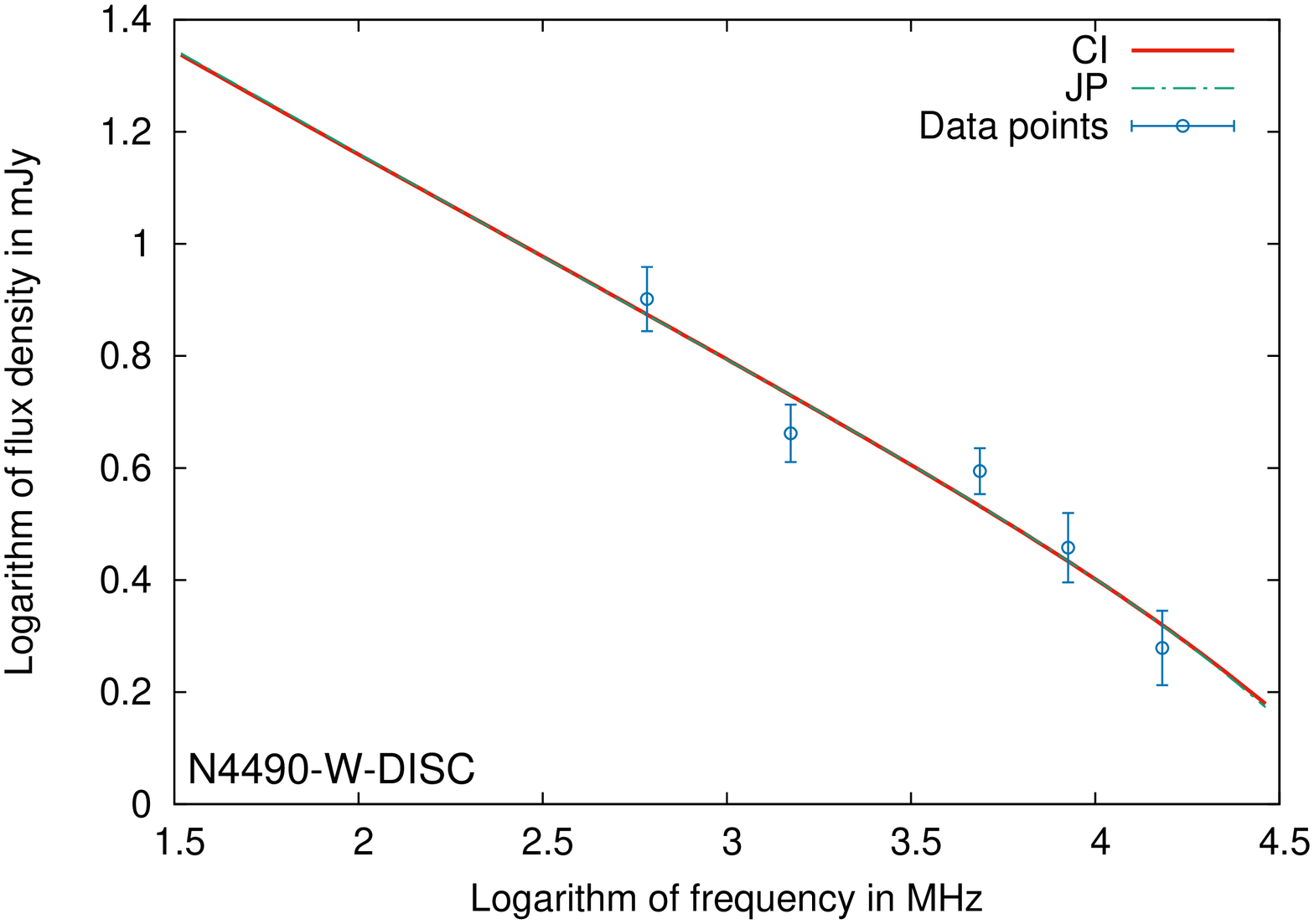}
    \end{subfigure}%
    \begin{subfigure}[p]{0.45\textwidth}
    \includegraphics[width=0.7\textwidth, angle=270]{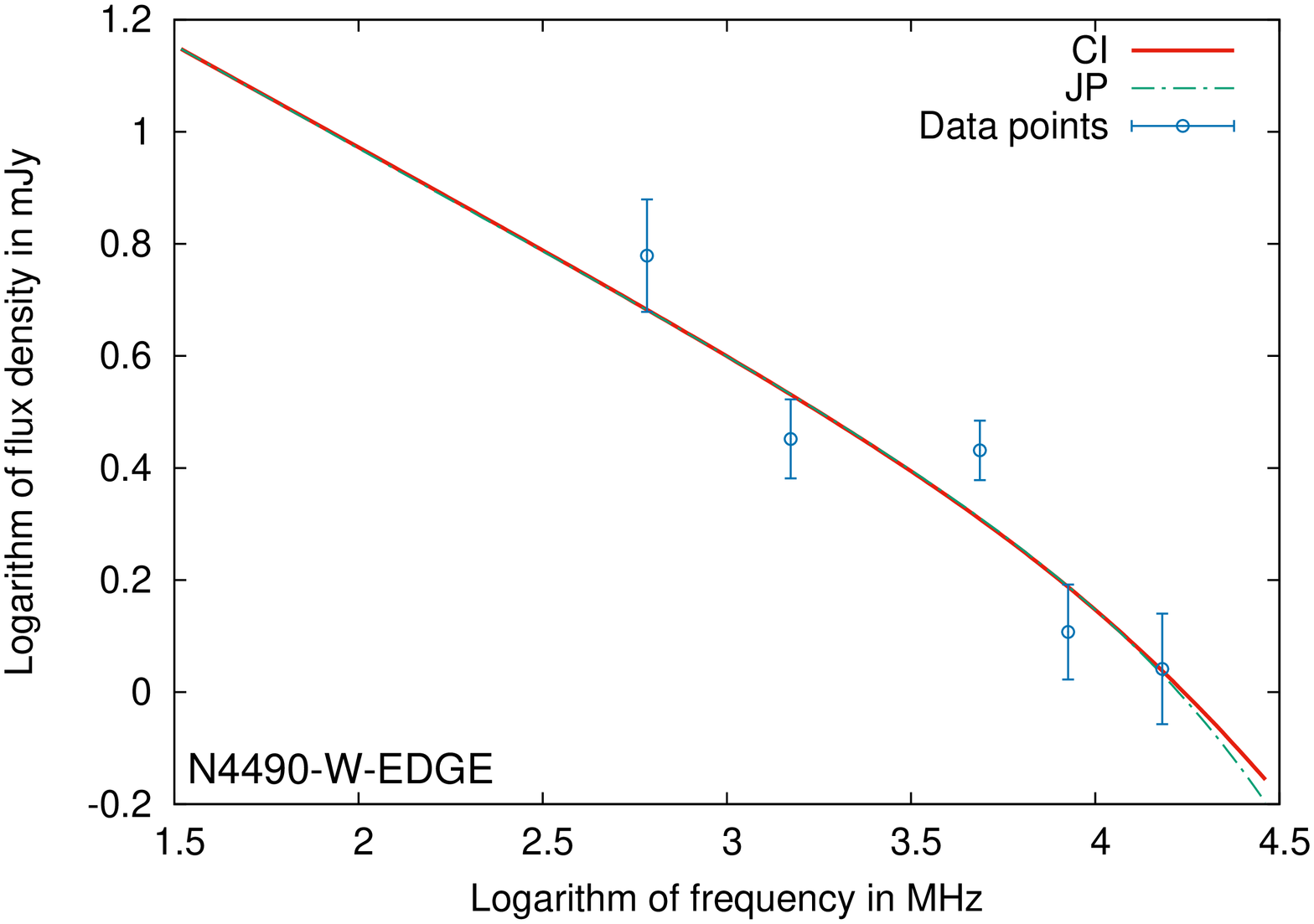}
      \end{subfigure}
    \caption{
	  Fitted Jaffe-Perola (dashed line) and Continuous Injection (solid line) model curves for the various regions of 
	  NGC\,4490\slash 85 galaxy pair. Explanation is provided in the text.
	}
    \label{fitimages}
\end{figure*}

A common assumption for all these three models
is that the magnetic field stays constant in time. The first two models 
present a similar approach to the problem, differing in the matter of the evolution
of the pitch angles of charged particles gyrating in the magnetic field.
JP assumes its isotropisation (time-scale of this process is
significantly lower than the radiative lifetime), whereas KP postulates its 
conservation. The third one (CI) is significantly different, as it includes 
the continuous injection of a power-law distributions of relativistic 
electrons. The resulting electron spectrum is then a combination of emission 
from various synchrotron populations. Final calculation of the spectral
age implies knowledge of the magnetic field strength. In the previous studies,
constant value of the magnetic field throughout the pair area was assumed.
To improve the age estimates, we used the values calculated in Sect.~\ref{magfield} 
(expressed in nT) and substituted them to the following equation:\\
\begin{equation}
\tau = 50.3 \frac{B^{1\slash 2}}{B^2 + B^{2}_{\mathrm{IC}}} \times (\nu_{\mathrm{break}}(1 + z))^{-1\slash 2} \mathrm{[Myr]}
\end{equation}
where $B_{\mathrm{IC}} = 0.338(1 + z)^{2}$ is the CMB magnetic field equivalent \citep{CMB}.
Obtained values of $\alpha_{\mathrm{inj}}$, $\nu_{\rm{break}}$, and of the spectral age are summarised in Table\,\ref{params},
while Fig.~\ref{fitimages} shows the results of spectral fitting (dependence of the non-thermal flux density density on the
frequency, both expressed in the logarithmic scale). The 22.46\,GHz data points were not taken into the account, as the flux density
at this frequency is much lower than expected, because of the diminished sensitivity of the interferometer.

\subsubsection{Constraints on the applicability and reliability of the age estimates}

Studies on the spectral age of an electron population always come with an element of uncertainty, as a model of the electron losses
-- which is an approximation to some, usually not precisely known, extent -- has to be assumed. In this study, we decided to use the 
JP and CI (which is basically a series of JP-based events) models, which have been originally designed to operate on rather compact
sources of electron acceleration. A question then arises, if they are applicable to a normal, non-AGN galaxy. Different reasons can be
provided for different entities described here. For the star-forming regions, the analogy is quite straight-forward and usage of the
JP and CI models comes without serious objections. In fact, it is a rather common assumption, and as \citet{torres} argues, for regions
of rapid star formation, the effect of diffusion is not essential. For a galaxy treated as a whole, the results from the aforementioned 
models become more tentative, as here diffusion -- especially at the outskirts of the disk -- becomes a significant (or the single one) 
source of the CRs. However, JP and CI models can still be useful for starburst galaxies and are still being used (eg. by \citealt{basu_model}). 
NGC\,4490 is a textbook example of a starburst galaxy, and the star formation rate is moreover constant throughout the disk 
\citep{clemens}, justifying usage of both JP and CI. The last studied structure, the intergalactic bridge seems a rather exotic target
for these two models. But yet, the electron population of this structure was in fact accelerated in a disk region of star formation, for
which both models have proven to be applicable. The observed spectrum can be treated as a one of a compact region of effective electron
supply, for which the acceleration has ceased at a given time.\\

Another issue is the zero spacing problem of the purely interefeormetric data. For compact and strong structures like the 
star-forming regions, or the galactic centre, it is not important. Owing to the rather constant star formation throughout the disk, its emission
should not be heavily affected. However, reliability of the spectral fits might be lower for the intergalactic bridge.
The final remark should address the common assumption for both models that the magnetic field remains constant in time, which is not true 
for the studied entities. However, the calculated values come with some uncertainty (of a few $\mu$G), which, in the first approximation, can
be treated as a term that \emph{expresses} the influence of the magnetic field evolution. This is especially true for the young, star-forming 
regions. Results for the disk should be treated with
respect to the limitation described above, whereas as the intergalactic bridge is a structure where the magnetic field is no longer 
amplified, over the course of time calculated from the modelling (and strength estimation uncertainties) its magnetic field can be 
regarded as moreover constant.
In some of the cases, the fitted spectra were nearly straight lines and the break frequency (crucial for the age estimate) was very weakly 
localised. A straight-line spectrum signifies a young source, and so are the star-forming regions. Thus, in such cases we decided to
present only an upper estimate of the spectral age.

\subsection{Coincidence between the radio and neutral hydrogen emission}
\label{hisect}

Among the most distinctive features of the NGC\,4490\slash 85 system are the giant neutral hydrogen tails that extend
north and south from NGC\,4490. The unusual {\rm H}{\sc i} morphology of this pair was first reported by \citet{dd},
who used the Jodrell Bank Mark II telescope to create square grids of pointings for a selected list of nearby galaxies. For 
NGC\,4490, they noticed that the gaseous halo is much larger in diameter than the optical counterpart. The first maps, made by
\citet{vab_hi}, revealed the morphology of the emission. Arp\,269 is a ``gas streamer'', a system in which the neutral gas 
is not confined between members (like it happens for eg. HCG\,95, \citealt{hcg95}, or NGC\,691, \citealt{n691}),
but extends beyond them. A good example is the Leo Triplet, \citealt{haynes}, or Arp\,143 -- \citealt{tail}). 
As the magnetic field is frozen into the gas, an issue if these intergalactic structures can transport magnetised matter outside 
the galaxy systems is still open, as the radio data on groups are scarce.
Such a hypothesis is a plausible one, as ambipolar diffusion could cause such behaviour.
If common, this mechanism can lead to the magnetisation of large volumes of intergalactic space -- a process
that likely takes place in the case of Arp\,244 (the Antennae galaxies, \citealt{antennae}). \\

\begin{figure}
\resizebox{\hsize}{!}{\includegraphics{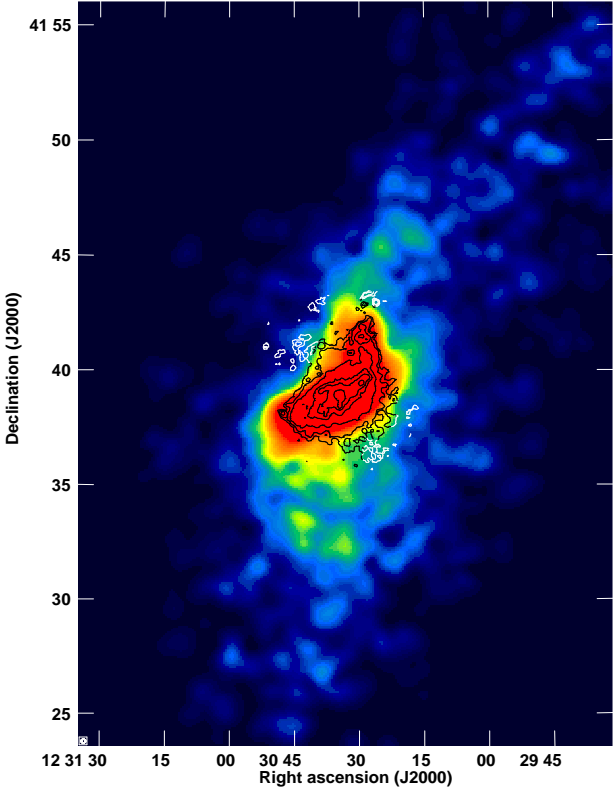}}
\caption{
 Contours of the 4.86\,GHz radio emission superimposed on the neutral hydrogen distribution map from \citet{whisp}, taken
 from the NED database. The contour levels are $3,5,10,20,30,50\times$ 33\,$\mu$Jy/beam 
 (first contour at the 2.5 times r.m.s. noise level).
 The angular resolution of the continuum data is  10\,arcsec. 
}
\label{hifig}
\end{figure}

In both of the aforementioned streamers, not even single spot of radio emission does coincide with the neutral hydrogen filaments
\citep{ourleo, our143}. Compared to them, NGC\,4490 looks promising, as -- so it can be seen in Fig.~\ref{hifig} -- both 
northern and southern extensions of the radio continuum envelope 
are embraced by the hydrogen one. However, the angle between the galactic disk and the magnetised outflow is different than the one
between the disk and neutral gas tails; whereas the gaseous tails are oriented in the NW--SE direction, the radio structures are oriented the 
other way round: NE--SW. As the continuum extensions do not exceed the innermost part of the {\rm H}{\sc i} halo, it is 
impossible to judge if the magnetic field is being pulled through these tails, or in another direction. Despite the question of 
the coincidence being left open, the morphology still suggests that a bubble of dense, magnetised matter does extend from the 
disk and might eventually be transported further as the matter propagates. As for the neutral component, it might be driven 
by the wind caused by supernovae having their progenitors in the still-active star forming regions of the galactic disk 
\citep{clemens_hi}, but not necessarily this is true for the magnetised matter. A study of the polarised emission
could be useful in studying this -- yet unexplained -- phenomenon.
Attempts to connect the continuum extension with the H$\alpha$ outflow detected by \citet{clemens_ram} were also unsuccessful,
as the radio extension reaches much further into the intergalactic space. \\

Such a doubt does not apply to the bridge that connects the pair members; in this case the intergalactic radio emission is 
accompanied by a stream of stars, and a maximum in the neutral gas emission. This clearly suggests the interaction-based origin
of this structure.

\subsection{Details on the galaxy pair and individual regions}
\label{sources}

\subsubsection{Galaxy pair as a whole}

The whole envelope of the radio emission is described well by both JP and CI models. Only small deviations from the model
curve can be seen (Fig.~\ref{fitimages}). The biggest outlier is the one that represents the 4.86\,GHz flux density. This is because of
the aforementioned extent of the radio envelope at this frequency. The injection spectral indices of 0.52
and 0.60 for JP and CI models, respectively are somewhat flatter than the non-thermal spectral index value
calculated by \citet{niklas28}, 0.76$\pm$0.03; however, this is not unusual, as the injection index refers to the value 
calculated at the lowermost edge of the SED, while the value given by \citet{niklas28} refers to the higher frequencies. 
As a non-thermal spectrum steepens with the increasing frequency due to the electron losses, these values can be considered 
to be in an agreement. Age estimates for the disk follow the general tendency for the CI model to return higher values. The JP model 
estimates the spectral age to be between 2.4 and 5.7\,Myrs, while CI returns 3.8 and 14.9\,Myrs, respectively. This suggests a 
rather young population of the electrons, which was still supplied in the nearest past of this galaxy.\\

As a supplement, we also performed fitting for the ``smooth disk'' -- the flux density from the galaxy pair with all the individual
regions of enhanced radio emission (hence, vigorous star formation) subtracted. These estimates are in 
general very similar to the ones derived for the galaxies as whole. From the flux density densities one can easily see that the diffuse
emission is dominating. The age estimates for the smooth disk are in general slightly lower, but these differences are
very small compared to the range between minimal and maximal age.\\

The total magnetic field strength is $21.9 \pm 2.9$\,$\mu$G. This suggests strong magnetic field in the whole system 
(similar values have been obtained for the smooth disk case), higher than the typical one found in spiral galaxies (9$\pm$
1.3\,$\mu$G, \citealt{niklas28}), and similar to that found in M51 \citep{fletcher}. It is remarkable, as NGC\,4490
is a rather small galaxy, compared to the latter one, and smaller systems are usually less efficient in magnetic field 
amplification \citep{beckbook}. The star formation rate (SFR), 
which is closely related to the magnetic field strengths, is very similar in NGC\,4490 and M\,51: \citet{clemens} estimates 
the SFR of NGC\,4490 to be about 4.7\,M$_{\odot}$\,yr$^{-1}$ (basing on the radio data), whereas \citet{calzetti_m51} suggests
4.3\,M$_{\odot}$\,yr$^{-1}$ for M\,51 (basing on the ultraviolet data, 50--100\,Myr old stellar populations). As 
NGC\,4490 is much smaller than the latter galaxy, the surface star formation rate is higher, providing a sensible explanation
for the estimated magnetic field strength.

\subsubsection{Intergalactic bridge}

NGC\,4490 and 4485 are connected by a narrow radio bridge, that coincides with a stream of stars. Surprisingly, there is not much
thermal flux density there, suggesting deficiency of star forming regions. This hypothesis is supported by the shape of the spectrum: it falls 
rather quickly and weakens abruptly already at 4.86\,MHz (where flux density decrement can not be administered to the zero spacing problem).
The injection spectral indices reveal flat initial spectrum, suggesting that the observed radio emission originated in regions of
effective supply of charged particles. The CI spectral age estimate for the intergalactic bridge (4.5 -- 16.9\,Myrs) is almost the 
same as for the galactic disk, whereas the JP estimates (3.7 -- 7.9\,Myrs) are the highest (among the other ones derived with the JP model)
for the structures in this 
system. For this region, the JP model provides a better fit to the data. The magnetic field of the bridge is strong, only 
slightly weaker than that found inside the galaxy. The density of the energy contained within is approximately 1.39 $\pm$ 0.28
$\mathrm{erg}\,\mathrm{cm^{-3}}$, of the same order as that between the Taffy galaxies \citep{taffy}. Given the above information, 
a possible conclusion is that the radio emission in the bridge is a product of magnetic field amplification inside one of the galaxies,
and the emitting matter has been torn from it during the interaction process (or due to the wind of star formation). As there are
no effective sources of accelerated particles (judging on the steepening of the spectrum), the spectral age
refers to the last acceleration event in the galactic disk, and thus it can be considered the starting point of the merging process.
As the extent of the bridge is equal to  few kiloparsecs, it would imply that the charged particles travel with velocities of several
kilometres per second, which is not an exotic value.

\subsubsection{Optical center of the galaxy}

The data points for the galaxy center show some scatter 
(Fig.~\ref{fitimages}), but the models fit well.
Estimated injection spectral index is flat (0.45 and 0.46 for JP and CI, respectively), close to
the theoretical minimum for the young supernovae remnants (0.3 -- \citealt{SNR}). The fitted models show 
a very weak curvature, and therefore estimation of the break frequency is complicated.
Derived values come with a large uncertainty, and the upper boundary 
(which corresponds to the upper age limit of 1.4\,Myrs for JP and 2.8\,Myrs for CI models) suggests extremely young 
electron population. This indicates ongoing star formation and electron supply. The magnetic field is strong (39.3 $\pm$ 
5.4\,$\mu$G), but not unusual for a central region of a nearby galaxy. \citet{beckbook} state that centers of barred spirals
are regions of intense star formation and strong magnetic fields can be found there; notable examples include NGC\,1097 with app.
55\,$\mu$G, or NGC\,1365 with a maximum of 63\,$\mu$G \citep{n1097}.

\subsubsection{Northern spiral arm}

The northern spiral arm can be easily distinguished from the galactic disk because of the three local emission maxima.
They show high fraction of thermal emission, which becomes dominant at frequencies higher than 4.86\,GHz. The remaining
non-thermal electron population is older than that of the central region. Here both lower and upper limits for the
spectral age can be derived. The several Myr wide range of possible values suggests moderately aged population of electrons. The 
magnetic field is strong (25.5$\pm$3.5\,$\mu$G), higher than in typical spiral galaxies, yet still comparable to the arm 
regions of M51.

\subsubsection{W-DISK}

This region lies in the western part of the disk of NGC\,4490. Compared to the arm regions, it hosts very similar magnetic
field with B$_{\rm TOT}$ = 26.0$\pm$3.5\,$\mu$G, but as the model curve (Fig.~\ref{fitimages}) is very weakly bent, break 
frequency is poorly determined and again, only the upper age limit can be derived. Values of 2.0\,Myrs (JP) and 3.7\,Myrs (CI) 
are reasonable for an area of intensive, ongoing star formation.

\subsubsection{W-EDGE}

The westernmost star forming region of the disk seems not to differ much from the other areas of enhanced star formation. It hosts
strong magnetic field (18.2$\pm$2.3\,$\mu$G), albeit somewhat weaker than in the neighbouring W-DISK region. For this SED 
(Fig.~\ref{fitimages}), the break frequency can not be estimated more accurately than for W-DISK, so only upper limit of the spectral age
can be estimated. Both models yield spectral ages of no more than several Myrs, once again reasonable for this kind of source.

\subsubsection{E-EDGE}

This unresolved source lies on the easternmost edge of the galactic disk of NGC\,4490. It shows barely any thermal emission,
suggesting that it is a background radio galaxy, not related to the foreground galactic system. Constant slope of the SED
(Fig.~\ref{fitimages}) causes poor localisation of the break frequency, but it is clearly visible that this source is a subject \
to a continuous electron supply, resulting in a low spectral age.

\subsubsection{S-SRC}

South from the centre of NGC\,4490, a point-like source, easily distinguishable from the surrounding emission (see Fig.~\ref{spixGC})
can be seen.
This source exhibits an extremely flat spectrum, with an average band-to-band (total) spectral index value of 0.13. Such a 
value clearly indicates an almost purely thermal character of the emission, as it agrees with the theoretical value of 0.1 
\citep{pacholczyk}. This conclusion is not astounding in the light of its identification as an H{\sc ii} region \citep{merlin}.
Radio emission from such a structure should be dominated  by the thermal radiation with the synchrotron  component almost
negligible. However, the thermal fraction derived from our LVL maps is much lower than expected, as it does not
exceed 40 per cent at any frequency; a value closer to 100 per cent would be expected. This is most likely due to the high
internal extinction in this dense area. High extinction of the thermal flux density has also been
noticed by \citet{merlin}, who suggested a possibility that the obscuring dust is related not to the region itself, but to the
surrounding atomic gas.

\subsubsection{Star forming regions of NGC\,4485}

The companion galaxy NGC\,4485 is a rather radio quiet one, compared to its neighbour. Most of its emission is due to a
group of three sources, compact in character. This group lies in the southern outskirts of the galactic disk, and  
\citet{clemens_ram} identifies it with a bow-shock in the IGM. Emission from these sources has a significant thermal component (reaching
50\% at 15.21\,GHz). Values for the remaining, non-thermal flux density lie on a more or less straight line, suggesting low spectral 
age. As in some of the previous cases, only upper limit for the spectral age can be derived: 7.8\,Myrs for the JP model and
23.5\,Myrs for the CI. These limits are higher than those obtained for the disk regions;
whereas this might indicate the older age of the structure formed by the shock, straight-line character
of the SED suggests that  this region is still a very young one. 
The magnetic field is somewhat weaker than that found in the intergalactic bridge. Lack of amplification 
suggests that the shock is rather weak and does not compress the magnetised IGM sufficiently to enhance the magnetic field bound to
it.

\subsection{Tracing the remnant of SN2008ax}
\label{snr}

On 3$^{{\rm rd}}$ March 2008, a supernova was discovered in the eastern part of NGC\,4490 disk, at $\mathrm{R.A._{2000}}=12^\mathrm{h} 
30^\mathrm{m} 40.8^\mathrm{s}$, $\mathrm{Dec_{2000}}=+41^{\circ} 38' 14.5''$ \citep{mostardi}. Denoted SN2008ax, it belongs to 
the rarest type of the SNe, namely Type IIb \citep{mostardi}: their progenitors are similar to those of the SNIb, but unlike 
the latter ones, they have small hydrogen envelopes \citep{woosley, pastorello}. There is no evidence for an increased radio flux density 
density at that position in our GMRT observations (Fig.~\ref{gband}); in the high-res maps used for spectral fitting, there is
no sign of emission at all. This means that the SNR flux density on the 1805th day since explosion does not exceed 4.07 $\pm$ 
0.06\,mJy (a mean value from a small area surrounding the position of SNR).

\subsection{Origin of the unusually steep spectrum of the distant galaxy}
\label{distgal}

A distant galaxy at $\mathrm{R.A._{2000}}=12^\mathrm{h}30^\mathrm{m} 26.1^\mathrm{s}$, $\mathrm{Dec_{2000}}=+41^{\circ} 42' 
58''$ (z = 0.124880, based on the SDSS data), not related to the Arp\,269 pair can be seen north from NGC\,4485 
in our 0.61, 1.49 and 4.86\,GHz maps (as a point source). 
This object also belongs to a galaxy system, as there is another galaxy at virtually the same redshift (z=0.124696),
located app. 140\,kpc away. 
Unfortunately, neither emission from the second component, nor from any intergalactic structure was detected.\\\\
Our analysis shows that this distant galaxy has a steep spectral index of 1.13 $\pm$ 0.07 between 
0.61 and 4.86\,GHz, suggesting domination of the non-thermal emission originating from a moderately aged population of 
relativistic electrons. As this object is 
identified as an AGN, much flatter spectrum might be expected. This distant galaxy may host a
Gigahertz-Peaked Source (GPS), or a Compact Steep Source (CSS, \citealt{CSS}). These sources are steep-spectrum, compact objects 
with the emission  peak around 100\,MHz (CSS) or 1\,GHz (GPS). Among possible explanations for their origin, two are being 
generally accepted as the most probable ones: either these are old, frustrated objects \citep{frust1, frust2}, obscured by
dust in the host galaxy, or young objects, that did not evolve sufficiently yet \citep{young}. Due to the significant distance 
(z = 0.125) the source is unresolvable at any of the frequencies used in this study. However, there is no evidence of radio 
structures that might extend beyond the contour of the optical emission. A quick look at the MIPS map shows 
that the dust emission locally peaks at the position of this distant galaxy. This might be a hint that the frustrated source 
hypothesis is indeed correct here.

\section{Conclusions} 
\label{conclusions}

We studied the nearby galaxy pair NGC\,4490\slash 85 (Arp\,269) at 0.61, 1.49, 4.86, 8.44, 14.93, 15.21 and 22.46\, GHz. The radio emission was 
separated into the thermal and non-thermal components. Magnetic field properties, CR electron synchrotron age and coincidence 
of the radio emission with the neutral gas distribution were examined. We came to the following conclusions:\\
\begin{enumerate}
 \item There is an intergalactic extension of the continuum emission envelope visible at several frequencies. This extension
 is double--sided, and is visible even in the higher frequency maps. 
 \item The aforementioned extension is oriented nearly perpendicular to the neutral hydrogen tail. The reasons for such a 
 configuration are yet unknown; it might suggest that there are indeed two different outflows, but the reasons for such 
 behaviour are yet unknown.
 \item The magnetic field strength -- the mean value for the galaxy ($21.9 \pm 2.9$\,$\mu$G), as well as values estimated for
 particular disk regions (18--40 $\mu$G) -- is high and comparable to the grand-design, or barred spiral galaxies, like M\,51,
 NGC\,1097, or NGC\,1365 -- despite NGC\,4490 being much smaller object. This can be explained on the basis of high surface 
 star formation rate, providing effective source of relativistic particles. 
  \item The age of the electron population of the disk has been studied using Jaffe-Perola and Continuous Injection models. No matter
 the model used, the resulting spectral ages are very low and suggest an ongoing star formation throughout both galaxies that 
 form the pair -- supporting the hypothesis that ongoing rapid star formation is responsible for the magnetic field strength.
  \item There is a magnetised bridge between member galaxies. It hosts strong magnetic field, and the age
  estimate -- which, for the JP model, is the highest obtained in the whole system -- suggests that the spectral age signifies the moment
  when merging process started (3.7--7.9\,Myrs and 4.5--16.9\,Myrs for the JP and CI models, respectively).
  \item There is no evidence of a supernovae remnant in the position of SN2008ax; the upper limit for this SN on the day 1805th 
 from explosion is therefore 4.07$\pm$0.06 mJy (mean flux density value in the closest vicinity of the suspected SNR's position).
 \item In the angular vicinity of NGC\,4490, a distant (z=0.125) galaxy with a very steep spectrum lies. This object might be a 
 compact steep source (CSS): an unevolved, or frustrated active galactic nuclei. This object forms a pair with another one, 
 although its companion is a radio-quiet one and there are no signs of intergalactic emission between them.
\end{enumerate}

\section*{acknowledgements}

We wish to thank the anonymous referee, whose comments and suggestions allowed us to 
significantly improve this article.\\
We are thankful to Dr Marcel Clemens for his 15.21\,GHz Ryle Telescope map that allowed us to improve our estimates based§
on the spectral properties of this galaxy system.\\
We thank the staff of the GMRT that made these observations possible. GMRT is run by the National Centre for Radio Astrophysics
of the Tata Institute of Fundamental Research.\\
This research has been supported by the scientific grant from the National Science Centre (NCN), dec. No.\,2011/03/B/ST9/01859.\\
This research has made use of the NASA/IPAC Extragalactic Database (NED) which is operated by the Jet Propulsion Laboratory,
California Institute of Technology, under contract with the National Aeronautics and Space Administration. 
This research has made use of NASA's Astrophysics Data System.
Funding for the SDSS and SDSS-II has been provided by the Alfred P. Sloan Foundation, the Participating Institutions, the 
National Science Foundation, the U.S. Department of Energy, the National Aeronautics and Space Administration, the Japanese 
Monbukagakusho, the Max Planck Society, and the Higher Education Funding Council for England. The SDSS Web Site is 
http://www.sdss.org/.


\label{lastpage}

\end{document}